\documentclass{article}
\usepackage{spconf,amsmath,amsthm,amssymb,graphicx}

\usepackage{pgf,tikz,pgfplots}
\usepackage{mathrsfs}
\usetikzlibrary{arrows}
\usepackage{mathtools} 
\usepackage{hyperref}
\hypersetup{
	colorlinks=true,
	linkcolor=blue,
	filecolor=magenta,
	urlcolor=cyan,
	citecolor=blue
}
\usepackage{subfigure}
\usepackage{ifthen}
\usepackage[utf8]{inputenc} 
\usepackage[T1]{fontenc}
\usepackage{caption}
\usepackage{algorithm,algorithmic}
\newtheorem{theorem}{\bf Theorem}
\usepackage{mathtools}
\usepackage{cuted}

\usepackage{cases}

\usepackage{float}

\newcommand{\LPRR}{LPRR}

\title{An efficient linear programming rounding-and-refinement algorithm for large-scale network slicing problem}
\name{Wei-Kun Chen$^{\star}$, Ya-Feng Liu$^{\dag}$\thanks{\noindent Ya-Feng Liu is the corresponding author.}, Yu-Hong Dai$^{\dag}$, and Zhi-Quan Luo$^\ddag$
}
\address{ 
	$^{\star}$School of Mathematics and Statistics, Beijing Institute of Technology, Beijing, China\\[2pt]
	$^{\dag}$LSEC, ICMSEC, AMSS, Chinese Academy of Sciences, Beijing, China\\[2pt]
	$^{\ddag}$Shenzhen Research Institute of Big Data and The Chinese University of Hong Kong, Shenzhen, China\\[2pt]
	Email:
	chenweikun@bit.edu.cn, \{yafliu,~dyh\}@lsec.cc.ac.cn, luozq@cuhk.edu.cn
}

\begin{document}
%
\ninept
\maketitle
\setlength{\abovedisplayskip}{0.14cm}
\setlength{\belowdisplayskip}{0.14cm}
\setlength{\jot}{0.14cm}
{
\begin{abstract}
In this paper, we consider the network slicing problem which attempts to map multiple customized virtual network requests (also called services) to a common shared network infrastructure and allocate network resources to meet diverse service requirements, and propose an efficient two-stage algorithm for solving this NP-hard problem.
In the first stage, the proposed algorithm uses an iterative linear programming (LP) rounding procedure to place the virtual network functions of all services into cloud nodes while taking traffic routing of all services into consideration;
in the second stage, the proposed algorithm uses an iterative LP refinement procedure to obtain a solution for traffic routing of all services with their end-to-end delay constraints being satisfied.
Compared with the existing algorithms which either have an exponential complexity or return a low-quality solution, our proposed algorithm achieves a better trade-off between solution quality and computational complexity.
In particular, the worst-case complexity of our proposed algorithm is polynomial, which makes it suitable for solving large-scale problems. 
Numerical results demonstrate the effectiveness and efficiency of our proposed algorithm.
\end{abstract}
\section{Introduction}
\label{sec:intro}

Network function virtualization (NFV) plays a crucial role in the fifth generation (5G) and beyond 5G networks \cite{Mijumbi2016}.
Different from traditional networks where service functions are processed by specialized hardwares in fixed locations, NFV efficiently takes the advantage of cloud technologies to configure some specific nodes (called cloud nodes) in the network to process network service functions on-demand, and then flexibly establishes a customized virtual network for each service request.
%
%
%
However, as virtual network functions (VNFs) of all services run over a shared common network infrastructure, it is crucial to allocate network (e.g., cloud and communication) resources to meet the diverse service requirements.

The above resource allocation problem in the NFV-enabled network is called \emph{network slicing} in the literature.
Various approaches have been proposed to solve it or its variants; see \cite{Chen2020}-\cite{Luizelli2015} and the references therein.
These approaches can generally be classified into two categories: (i) exact algorithms that solve the problem to global optimality and (ii) heuristic algorithms that aim to quickly find a feasible solution for the problem.
In particular, references \cite{Chen2020}-\cite{Addis2015} proposed the mixed integer linear programming (MILP) formulations for the network slicing problem and used standard MILP solvers like Gurobi \cite{Gurobi} to solve their problem formulations.
References \cite{Hu2013}-\cite{Liu2017} proposed a column generation approach \cite{Conforti2014} to solve the related problems.
Though the above two approaches can solve the network slicing problem to global optimality, they generally suffer from low computational efficiency as their worst-case complexities are exponential.
Due to this, references \cite{Zhang2017}-\cite{Luizelli2015} simplified the solution approach by decomposing the network slicing problem into a VNF placement  subproblem (which maps VNFs into cloud nodes in the network) and a traffic routing subproblem (which finds paths connecting two adjacent VNFs in the network) and solving each subproblem separately.
%
To obtain a binary solution for the VNF placement subproblem, references \cite{Zhang2017,Chowdhury2012} first solved the linear programming (LP) relaxation of the network slicing problem and then used a rounding strategy while references \cite{Lischka2009}-\cite{Luizelli2015} used some greedy heuristics (without solving any LP).
Once the VNFs are mapped to the cloud nodes, the traffic routing subproblem is solved by using shortest path, $k$-shortest path, or multicommodity flow algorithms.
However, solving the VNF placement subproblem without taking the global information (i.e., traffic routing of all services) into account can lead to infeasibility or low-quality solutions.
Therefore, algorithms that find a high-quality solution of the network slicing problem while still enjoy a polynomial-time complexity are still highly needed.

In this paper, we propose a two-stage LP rounding-and-refinement algorithm which achieves a good trade-off between high solution quality and low computational complexity.
%
%
Specifically, in the first stage, we solve the VNF placement subproblem by using an iterative LP rounding procedure, which takes traffic routing into account;
in the second stage, we solve the traffic routing subproblem by using an iterative LP refinement procedure to find a solution that satisfies the end-to-end (E2E) delay constraints of all services.
In particular, the proposed algorithm has a guaranteed polynomial-time worst-case complexity, and thus is particularly suitable for solving large-scale problems.
Numerical results demonstrate the effectiveness and efficiency of our proposed algorithm over the existing ones.

\section{System model and problem formulation}
\label{sec:modelformulation}
Let $\mathcal{G}=\{\mathcal{I},\mathcal{L}\}$ be the  directed network, where $\mathcal{I}=\{i\}$ and $\mathcal{L}=\{(i,j)\}$ are the sets of nodes and links, respectively. 
Each link $ (i,j) $ has an expected (communication) delay $ d_{ij} $ \cite{Woldeyohannes2018,Mohammadkhan2015,Luizelli2015}, and a total data rate upper bounded by the capacity $C_{ij}$.
The set of cloud nodes is denoted as $ \mathcal{V} \subseteq \mathcal{I}$.
Each cloud node $ v $ has a computational capacity $ \mu_v $ and processing one unit of data rate requires one unit of (normalized) computational capacity, as assumed in \cite{Zhang2017}.
A set of flows $\mathcal{K}=\{k\}$ is required to be supported by the network.
The source and destination nodes of flow $k$ are denoted as $S(k)$ and $D(k)$, respectively, with $S(k),D(k)\notin \mathcal{V}$.
Each flow $ k $ relates to a customized service, which is given by a service function chain (SFC) consisting of $ \ell_k $ service functions that have to be processed in sequence by the network: $f_{1}^k\rightarrow f_{2}^k\rightarrow \cdots \rightarrow f_{\ell_k}^k$ \cite{Zhang2013,Halpern2015,Mirjalily2018}.
%
To minimize the coordination overhead, each function must be processed at exactly one cloud node, as required in \cite{Domenico2019,Zhang2017,Woldeyohannes2018}.
If function $ f^k_s $, $ s \in \mathcal{F}(k) := \{1,\ldots, \ell_k\} $, is processed by
cloud node $ v $ in $ \mathcal{V} $, the expected NFV delay is assumed to be known as $
d_{v,s}(k) $, which includes both processing and queuing delays \cite{Woldeyohannes2018,Luizelli2015}.
For flow $ k $, the service function rates before receiving any function and after receiving {function} $ f^k_s $ are denoted as $ \lambda_0(k) $ and $ \lambda_s(k) $, respectively.
Each flow $ k $ has an E2E delay requirement, denoted as $ \Theta_k $.
%
%
%
%

The network slicing problem is to determine functional instantiation, the routes, and the associated data rates on the corresponding routes of all flows while satisfying the capacity constraints on all cloud nodes and links, the SFC requirements, and the E2E delay requirements of all flows. 
Next, we shall briefly introduce the problem formulation; see more details in \cite{Chen2020}.  \vspace{0.1cm}\\
{\bf\noindent$\bullet$ VNF Placement\vspace{0.1cm}\\}
\indent We introduce the binary variable $x_{v,s}(k)$ to indicate whether or not function $f^k_s$ is processed by cloud node $v$.
%
%
%
Each function $f_s^k$ must be processed by exactly one cloud node, i.e.,
\begin{eqnarray}
\label{onlyonenode}
\sum_{v\in \mathcal{V}}x_{v,s}(k)=1,~\forall ~k \in \mathcal{K},~ \forall ~s\in  \mathcal{F}(k).
\end{eqnarray}
%
Let $y_v=1$ denote that cloud node $v$ is activated and powered on; otherwise $y_v=0$. Thus
\begin{equation}
\label{xyrelation}
x_{v,s}(k) \leq  y_v, ~ \forall~v \in \mathcal{V},~\forall~k \in \mathcal{K},~\forall~s \in \mathcal{F}(k). 
\end{equation}
The node capacity constraints can be written as follows:
\begin{equation}
\label{nodecapcons}
\sum_{k\in \mathcal{K}}\sum_{s \in \mathcal{F}(k)}\lambda_s(k)x_{v,s}(k)\leq \mu_v y_v,~\forall~ v \in \mathcal{V}.
\end{equation}
{\bf\noindent$\bullet$ Traffic Routing\vspace{0.1cm}\\}
\indent Let $ (k,s) $ denote the flow which is routed between the two cloud nodes hosting two adjacent functions $ f_s^k $ and $ f_{s+1}^k $.
%
Similar to \cite{Chen2020}, we suppose that there are at most $P$ paths that can be used to route flow $(k,s)$ and denote $\mathcal{P}=\{1, \ldots, P\}$.
%
Let $ r(k,s,p) $ be the fraction of data rate $\lambda_s^k$ on the $ p $-th path of flow $ (k,s) $.
%
Then, the following constraint enforces that the total data rate between the two nodes hosting functions $ f_s^k $ and  $ f_{s+1}^k $ is equal to $ \lambda_s(k) $:
\begin{align}
& \sum_{p \in \mathcal{P}}  r(k, {s}, p) =  1,   ~ \forall~ k \in \mathcal{K},~\forall~s\in \mathcal{F}(k)\cup \{0\} \label{relalambdaandx11}.                                  
\end{align}
Let $ z_{ij}(k,s,p)\in \{0,1\}$ denote whether or not link $ (i,j) $ is on the $ p $-th path of flow $ (k,s) $ and $ r_{ij}(k,s,p) $ be the associated fraction of data rate $\lambda_{s}^k$. Then
\begin{align}
& r_{ij}(k, s,  p ) = r(k,s,p) z_{ij}(k, s,p ), \nonumber                                                                                          \\
& ~~\forall~(i,j) \in {\mathcal{L}}, ~\forall~k \in \mathcal{K}, ~\forall~s \in \mathcal{F}(k)\cup \{0\},~\forall~p \in \mathcal{P}. \label{nonlinearcons}
\end{align}
The total data rates on link $ (i,j) $ is upper bounded by capacity $ C_{ij} $:
\begin{equation}
\label{linkcapcons1}
\sum_{k \in \mathcal{K}} \sum_{s\in \mathcal{F}(k) \cup \{0\}}\sum_{p \in \mathcal{P}} \lambda_{s}(k) r_{ij}(k, s,p) \leq C_{ij}, ~  \forall~(i,j) \in \mathcal{L} .
\end{equation}
%
%
%
%
%
%

To ensure that the functions of each flow $k$ are processed in the prespecified order $f_{1}^k\rightarrow f_{2}^k\rightarrow \cdots \rightarrow f_{\ell_k}^k$ and for each $s \in \mathcal{F}(k)\cup \{0\}$ and $p \in \mathcal{P}$, $\{(i,j) : z_{ij}(k,s,p)=1 \}$ forms a path, we need the flow conservation constraint \eqref{mediacons2}.
\begin{figure*}[t]
	\begin{equation}
		\sum_{j: (j,i) \in \mathcal{{L}}} z_{ji}(k, s, p) - \sum_{j: (i,j) \in \mathcal{{L}}} z_{ij}(k, s,  p)=\left\{\begin{array}{ll}
		0,     & \text{if}~ i   \in {\mathcal{I}}\backslash {\mathcal{V}};  \\
		x_{i,s+1}(k)- x_{i,s}(k),     & \text{if}~  i \in {\mathcal{V}},     
		\end{array} \right.	\forall~k \in \mathcal{K},~\forall~s \in \mathcal{F}(k)\cup \{0\},~ \forall~ p \in \mathcal{P} \label{mediacons2}.
		\end{equation}
		\vspace{-7mm}
\end{figure*}
%
%

Let $ \theta(k,s) $ denote the communication delay due to the traffic flow from the cloud node hosting function $ f^k_s $ to the cloud node hosting function $ f^k_{s+1} $. Then
%
%
\begingroup
\allowdisplaybreaks
\begin{align}
& \theta(k,s) \geq \sum_{(i,j) \in \mathcal{{L}}}  d_{ij}  z_{ij}(k, s, p),\nonumber                                                   \\
& \qquad\qquad  \forall~k \in \mathcal{K}, ~ \forall~s \in \mathcal{F}(k) \cup \{0\},~\forall ~p \in \mathcal{P} \label{consdelay2funs1}.
\end{align}
\endgroup
%
%
%
To ensure that flow $k$'s {E2E} delay is less than or equal to its threshold $\Theta_k$, we need the following constraint:
\begin{equation}
\label{delayconstraint}
\theta_N(k) +\theta_L(k)  \leq \Theta_k,~\forall~k \in  \mathcal{K},
\end{equation}
where  $\theta_N(k) =  \sum_{v \in \mathcal{{V}}}\sum_{s \in \mathcal{F}(k)} d_{v,s}(k) x_{v,s}(k)$ and $\theta_L(k) = \sum_{s \in \mathcal{F}(k)\cup \{0\}} \theta(k,s)$ are the  total NFV delay on the nodes and the total communication delay on the links of flow $ k $, respectively.
\vspace{0.1cm}\\
{\bf\noindent$\bullet$ Problem Formulation\vspace{0.1cm}\\}
\indent The network slicing problem is to minimize a weighted sum of the total power consumption of the whole cloud network and the total delay of all services:
\begin{align}
& \min_{\boldsymbol{x},\boldsymbol{y},\boldsymbol{r},\boldsymbol{z},\boldsymbol{\theta}} &  & \sum_{v \in \mathcal{V}}y_v + \sigma \sum_{k \in \mathcal{K}} (\theta_L(k) + \theta_N(k)) \nonumber \\
& ~~~~~{\text{s.t.~}}                                                                    &  & (\ref{onlyonenode})-(\ref{delayconstraint}), \nonumber \\
& &&  x_{v,s}(k),~y_v\in\{0,1\},\,\forall \,k\in\mathcal{K}, ~s\in \mathcal{F}(k),~v\in\mathcal{{V}}, \nonumber\\
& && r(k,s,p),~r_{ij}(k, s, p )\geq 0,~z_{ij}(k, s, p )\in \{0,1\}, \nonumber\\
& && \qquad \forall~(i,j)\in \mathcal{L},~k\in \mathcal{K},~s\in \mathcal{F}(k)\cup \{0\},~p \in \mathcal{P}, \nonumber\\
& &&  \theta(k,s)\geq 0,~\forall~k \in \mathcal{K}, s \in \mathcal{F}(k)\cup \{0\},
\label{mip}
\tag{\rm{NS}}
\end{align}
where $ \sigma $ is a constant value that balances the two terms in the objective function.
It has been shown in \cite{Chen2020} that problem \eqref{mip} can be equivalently reformulated as an MILP problem and thus can be solved using standard MILP solvers like Gurobi.

The following Theorem \ref{NPhardness} shows the (strong) NP-hardness of problem \eqref{mip} in two very special cases and thus reveals the intrinsic difficulty of solving it.
This motivates us to develop efficient algorithms for approximately solving problem \eqref{mip}, especially when the problem's dimension is large.
\begin{theorem}
	\label{NPhardness}
	(i) Problem \eqref{mip} is NP-hard even when there is only a single service.
	(ii) Problem \eqref{mip} is strongly NP-hard even when each node's capacity, link's capacity, and service's E2E delay threshold are infinity.
	Moreover, there does not exist a constant approximation algorithm to solve it in this case.
\end{theorem}

\section{An LP rounding-and-refinement algorithm}
In this section, we focus on designing an efficient algorithm to obtain a high-quality solution for problem \eqref{mip}.
To do this, we first derive a compact LP relaxation for the problem and then develop a two-stage LP rounding-and-refinement algorithm based on it.
The basic idea of the proposed algorithm is to decompose the hard problem \eqref{mip} into two relatively easy subproblems and solve two subproblems separately \emph{while taking their connection into account}. 
Specifically, in the first stage, we find a binary vector $(\bar{\boldsymbol{x}},\bar{\boldsymbol{y}})$ for the VNF placement subproblem (i.e., $(\bar{\boldsymbol{x}},\bar{\boldsymbol{y}})$ satisfying constraints \eqref{onlyonenode}-\eqref{nodecapcons}) using an iterative LP rounding procedure, which takes traffic routing into account.
In the second stage, based on the binary vector $(\bar{\boldsymbol{x}},\bar{\boldsymbol{y}})$, we use an LP refinement procedure to solve the traffic routing subproblem to obtain a solution that satisfies the E2E delay constraints \eqref{consdelay2funs1}-\eqref{delayconstraint} of all services.
\vspace{0.1cm}\\
%
{\bf\noindent$\bullet$ A Compact LP Relaxation\vspace{0.1cm}
}

As problem \eqref{mip} can be reformulated as an MILP problem \cite{Chen2020}, simply relaxing the binary variables $\{y_v\}$, $\{x_{v,s}(k)\}$, and $\{z_{ij}(k,s,p)\}$ to be continuous variables will give a natural LP relaxation.
Recall that in problem \eqref{mip}, in order to model different paths for flow $(k,s)$, we introduce the notation $\{ p  :  p \in \mathcal{P} \}$ and use $\{(i,j) :r_{ij}(k,s,p)>0\}$ to represent the $p$-th path of flow $(k,s)$ (cf. \eqref{nonlinearcons} and \eqref{mediacons2}).
However, as $z_{ij}(k,s,p) \in [0,1]$ in the above natural LP relaxation, the traffic flow $\{(i,j) :r_{ij}(k,s,p)>0\}$ can be split into multiple paths. 
%
%
This reveals that there is some redundancy in the natural LP relaxation, i.e., we do not need to introduce the notation $\{p: p\in \mathcal{P}\}$ to model different paths for flow $(k,s)$ in it.

Inspired by this observation, below we derive a compact LP relaxation for problem \eqref{mip}.
Our strategy is to simply set $\mathcal{P}=\{1\}$.
Then by \eqref{relalambdaandx11}, we have $r(k,s,1) = 1$, and hence constraint \eqref{nonlinearcons} reduces to $r_{ij}(k,s,1)=z_{ij}(k,s,1)$.
Furthermore, we can remove constraints \eqref{relalambdaandx11}, \eqref{nonlinearcons}, and variables $\{r_{ij}(k,s,p)\}$, and replace constraint \eqref{linkcapcons1} by
\begin{equation}
\label{linkcapcons2}
\tag{\rm{6}'}
\sum_{k \in \mathcal{K}} \sum_{s\in \mathcal{F}(k) \cup \{0\}} \lambda_{s}(k) z_{ij}(k, s,1) \leq C_{ij}, ~  \forall~(i,j) \in \mathcal{L}.
\end{equation}
The natural LP relaxation then reduces to
\begin{align}
& \min_{\boldsymbol{x},\boldsymbol{y},\boldsymbol{z},\boldsymbol{\theta}} &  & \sum_{v \in \mathcal{V}}y_v + \sigma \sum_{k \in \mathcal{K}} (\theta_L(k) + \theta_N(k)) \nonumber \\
& ~~~~~{\text{s.t.~}}                                                                    &  & \eqref{onlyonenode}-\eqref{nodecapcons}, \eqref{linkcapcons2}, \eqref{mediacons2}-\eqref{delayconstraint}, \nonumber \\
& &&  x_{v,s}(k),~y_v\in[0,1],\,\forall \,v\in\mathcal{{V}},~k\in\mathcal{K}, s\in \mathcal{F}(k), \nonumber\\
& && z_{ij}(k, s,1 )\in [0,1], ~\theta(k,s)\geq 0,\nonumber \\
& & & \qquad \forall~(i,j)\in \mathcal{{L}}, ~k\in \mathcal{K},~s\in \mathcal{F}(k)\cup \{0\}. 
\label{lp}
\tag{\rm{NS-LP}}
\end{align}
\begin{theorem}
	\label{LPrelaxation}
	The LP problem \eqref{lp} is a relaxation of problem \eqref{mip} with any ${P} \geq 1$.
\end{theorem}
Theorem \ref{LPrelaxation} shows that the above LP problem \eqref{lp} is also a relaxation of problem \eqref{mip}.
Note that the numbers of variables and constraints in problem \eqref{lp} are much smaller than those in the natural LP relaxation of problem \eqref{mip}, especially when $P$ is large.
As a result, solving problem \eqref{lp} should be much more efficient than solving the natural LP relaxation.\vspace{0.1cm}\\
{\bf\noindent$\bullet$ Solving the VNF Placement Subproblem\vspace{0.1cm}}

Next, we solve the VNF placement subproblem by constructing a binary vector $(\bar{\boldsymbol{x}}, \bar{\boldsymbol{y}})$ that satisfies constraints \eqref{onlyonenode}-\eqref{nodecapcons}.
Since vector $\bar{\boldsymbol{y}}$ can be uniquely determined by vector $\bar{\boldsymbol{x}}$, in the following we concentrate on constructing the binary vector $\bar{\boldsymbol{x}}$.
To do this, we first solve the LP relaxation problem  \eqref{lp}, denoted its solution by $(\boldsymbol{x}^*, \boldsymbol{y}^*, \boldsymbol{z}^*,\boldsymbol{\theta}^*)$.
If $\boldsymbol{x}^*$ is a binary vector, we obtain a feasible solution $\bar{\boldsymbol{x}}:=\boldsymbol{x}^*$ for the VNF placement subproblem.
Otherwise, we set $x_{v,s}(k)=1$ in problem \eqref{lp} if $x^*_{v,s}(k)=1$.
Then we choose one variable, denoted as  $x_{v_0,s_0}(k_0)$, whose value $x^*_{v_0,s_0}(k_0)$ is the largest among the remaining variables, i.e.,  
\begin{align}
x^*_{v_0,s_0}(k_0)=  \max \big \{  x^*_{v,s}(k) \, : \, 0 < x^*_{v,s}(k)< 1, ~ v \in \mathcal{V}, \nonumber \\
\qquad k\in \mathcal{K},~s\in \mathcal{F}(k)\big\}. \label{maxvalue}
\end{align}
Next we decide to round variable $x_{v_0,s_0}(k_0)$ to one or zero.
In particular, we first set $x_{v_0,s_0}(k_0)=1$ in problem \eqref{lp}.
If the modified LP is infeasible, we set $x_{v_0, s_0}(k_0) =0$ and continue to round other variables respect to the values $\{x^*_{v,s}(k)\}$.
Otherwise, the modified LP is feasible and we repeat the above procedure to the solution of the modified LP until a binary solution is obtained.
The details are summarized in the following Algorithm \ref{roundingx}.
\begin{figure*}[!h]
	\centering
	\subfigure[]{
		\includegraphics[height=1.75in]{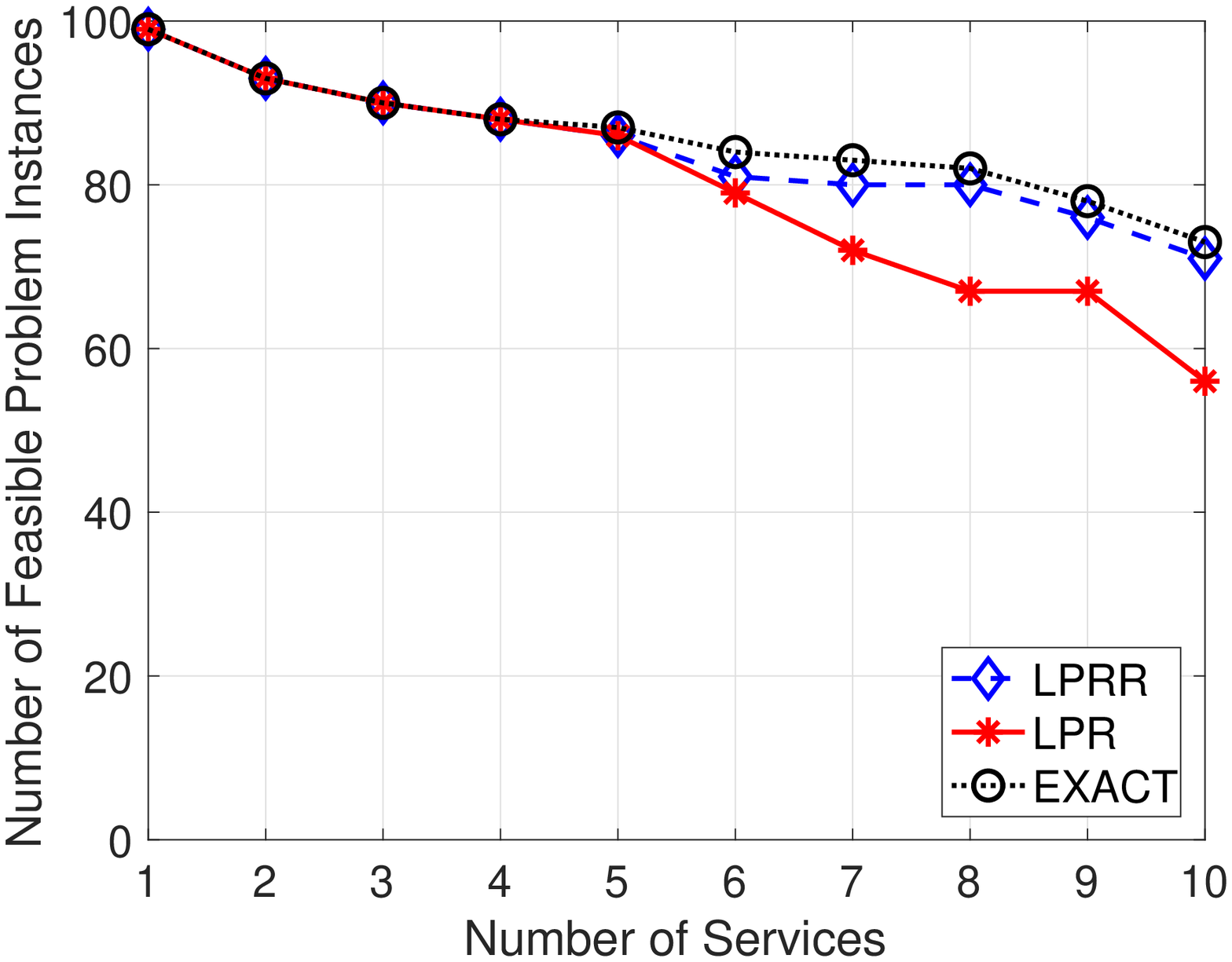}
		\label{nfeas}
	}
	\subfigure[]{
		\includegraphics[height=1.75in]{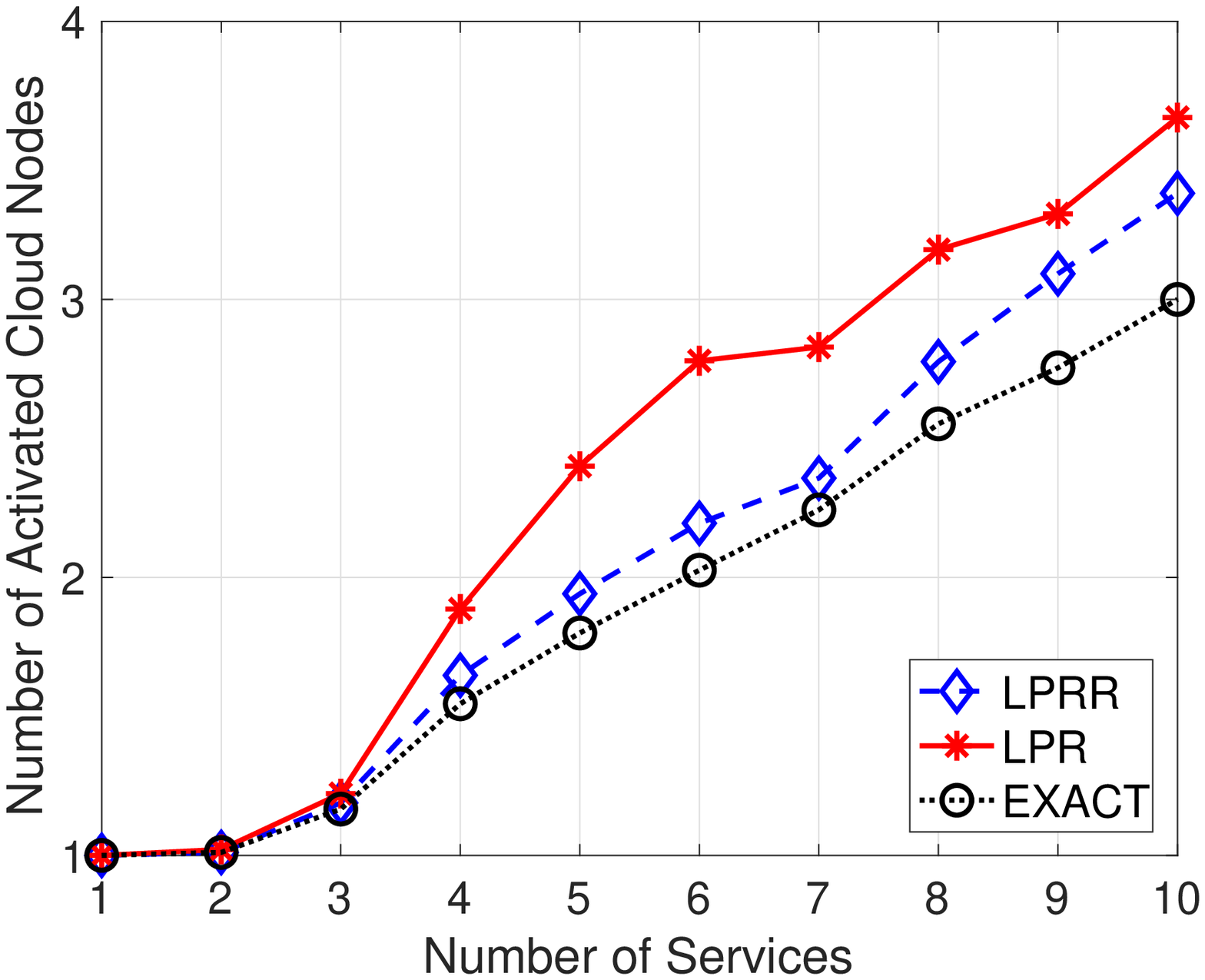}
		\label{sumy}
	}
	\subfigure[]{
		\includegraphics[height=1.75in]{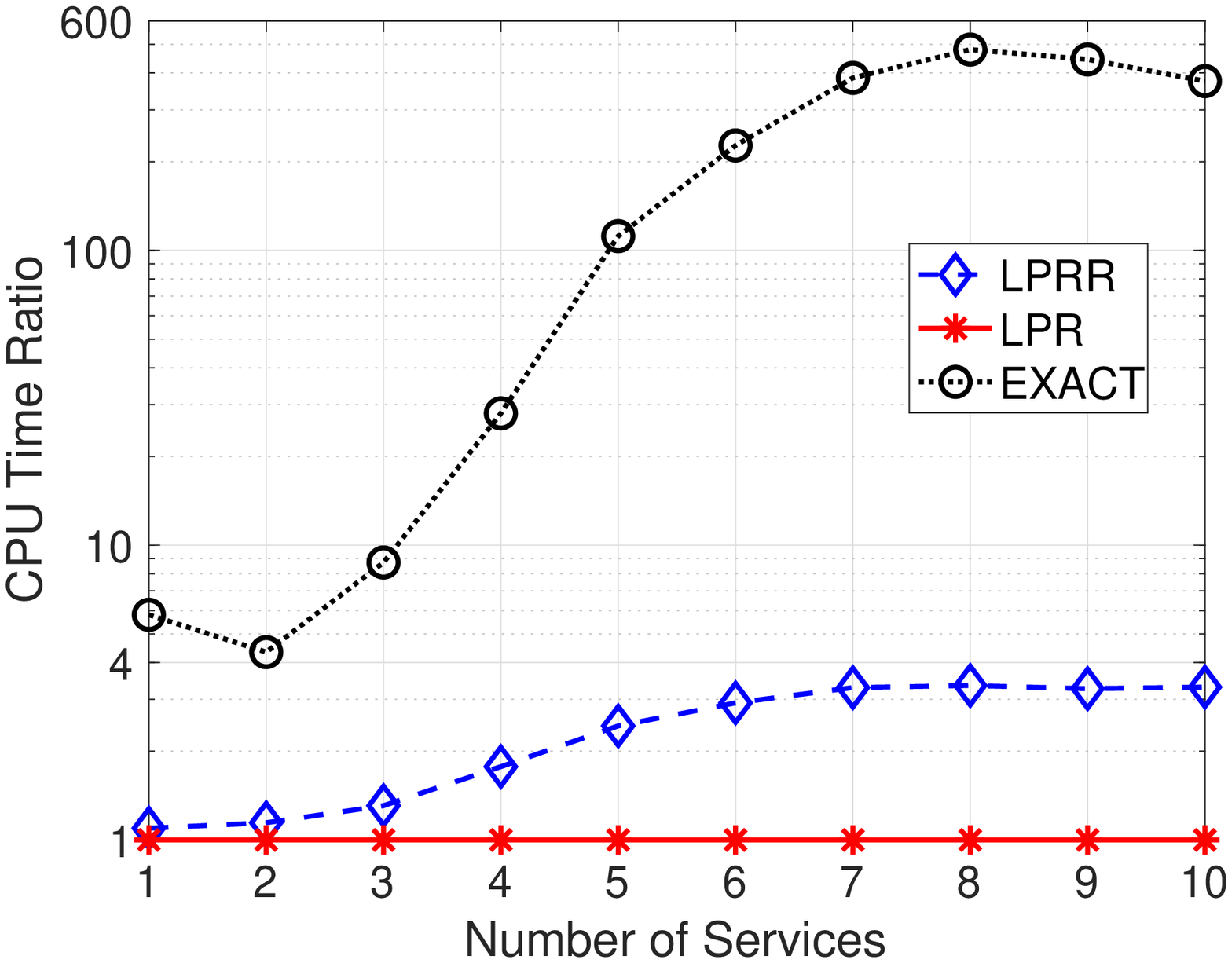}
		\label{timeratio}
	}
	\caption{{Performance comparison of \LPRR, LPR, and EXACT: (a) Number of feasible problem instances; (b) Average number of activated cloud nodes; (c) Average CPU time ratio.}}
	\vspace{-3.5mm}
	\label{figure1}
\end{figure*}
\begin{algorithm}[!h]
	\caption{{An iterative LP rounding procedure for solving the VNF placement subproblem}}
	\label{roundingx}
	\begin{algorithmic}[1]
		{
			\renewcommand{\algorithmicrequire}{\textbf{Input:}}
			\renewcommand{\algorithmicensure}{\textbf{Output:}}
			\STATE Initialize the set $\mathcal{A}=\varnothing$;
			\STATE Solve problem \eqref{lp} to obtain its solution $(\boldsymbol{x}^*,\boldsymbol{y}^*,\boldsymbol{z}^*,\boldsymbol{\theta}^*)$;
			\WHILE {(there exists some $v \in \mathcal{V}$, $k\in \mathcal{K}$, and $s\in \mathcal{F}(k)$ such that $0 < x^*_{v,s}(k) < 1 $)}
			\STATE For each $v \in \mathcal{V}$, $k\in \mathcal{K}$, and $s\in \mathcal{F}(k)$ with $x^*_{v,s}(k)=1$, if constraint $x_{v,s}(k)=1$ is not in set $\mathcal{A}$, add it into set $\mathcal{A}$;
			\STATE Let $(v_0, s_0,k_0)$ be the tuple in \eqref{maxvalue}. Add constraint $x_{v_0,s_0}(k_0)=1$ into set $\mathcal{A}$;
			\STATE Add the constraints in set $\mathcal{A}$ into problem \eqref{lp} to obtain a modified LP;
			\STATE If the modifed LP is feasible, let $(\boldsymbol{x}^*,\boldsymbol{y}^*,\boldsymbol{r}^*,\boldsymbol{\theta}^*) $ be its solution; otherwise, replace constraint $x_{v_0,s_0}(k_0)=1$ by constraint $x_{v_0,s_0}(k_0)=0$ in set $\mathcal{A}$ and set  $x^*_{v_0,s_0}(k_0)\leftarrow 0 $;
			\ENDWHILE
			\STATE If vector $(\boldsymbol{x}^*,\boldsymbol{y}^*)$ satisfies constraints \eqref{onlyonenode}-\eqref{nodecapcons}, then the binary vector $(\bar{\boldsymbol{x}},\bar{\boldsymbol{y}})\leftarrow(\boldsymbol{x}^*,\boldsymbol{y}^*)$ is feasible for the VNF placement subproblem; otherwise declare that the algorithm fails to find a feasible solution.
		}
	\end{algorithmic} 
\end{algorithm}

The above rounding strategy makes sure that we can round one variable, taking a fractional value at the current solution, at a time and more importantly this variable can be rounded to a binary value that is consistent to other already rounded variables.
This is in sharp contrast to the algorithm in \cite{Chowdhury2012} where the variables are rounded without ensuring the consistency of the current rounding variable with other already rounded variables.
%
It is worth remarking that our rounding strategy takes traffic routing into account (as the modified LP contains the information of traffic routing of all services).\vspace{0.1cm}\\
{\bf\noindent$\bullet$ Solving the Traffic Routing Subproblem\vspace{0.1cm}\\}
\indent Once we get a binary vector $(\bar{\boldsymbol{x}}, \bar{\boldsymbol{y}})$, we still need to solve the traffic routing subproblem by fixing $\boldsymbol{x}=\bar{\boldsymbol{x}}$ and $\boldsymbol{y}=\bar{\boldsymbol{y}}$ in problem \eqref{mip}.
In this case, the objective function in problem \eqref{mip} reduces to $g(\boldsymbol{\theta})=\sum_{k \in\mathcal{K}} \theta_L(k)$.
Similarly, we solve the LP problem \eqref{lp} with $\boldsymbol{x}=\bar{\boldsymbol{x}}$ and $\boldsymbol{y}=\bar{\boldsymbol{y}}$ to obtain a solution $(\boldsymbol{z}^*, \boldsymbol{\theta}^*)$.
Due to the (possible) fractional values of $\{ z^*_{ij}(k,s,1) \}$, $\theta^*(k,s)$ can be larger than the communication delay incurred by the traffic flow from the node hosting function $f_s^k$ to the node hosting function $f_{s+1}^{k}$.
To recompute the communication delay based on solution $(\boldsymbol{z}^*, \boldsymbol{\theta}^*)$, we need to solve the NP-hard {Min-Max-Delay} problem \cite{Liu2017a}.
Fortunately, there exists an efficient polynomial-time ($1+\epsilon$)-approximation algorithm for this problem \cite{Liu2017a}.
%
%
After recomputing the communication delays between all pairs of nodes hosting two adjacent functions, we can compute the total delay of each service $k$, denoted as $\bar{\theta}(k)$.
If $\bar{\theta}(k)> \Theta(k)$ for some service $k$, the current routing strategy is infeasible as it violates the E2E delay constraint of service $k$.  
We then use an iterative LP refinement procedure to try to get a solution that satisfies the E2E delay constraints of all services.

The idea of our refinement procedure is to increase the weights of the variables $\theta_{L}(k) $ corresponding to the service whose E2E delay constraint is not satisfied at the current solution, in order to refine the solution.
In particular, we change the objective function $g(\boldsymbol{\theta})$ in problem \eqref{lp} into $\hat{g}(\boldsymbol{\theta})=\sum_{k \in \mathcal{K}} \omega_k \theta_L(k)$ where $\omega_k \geq 1 $ for all $k \in \mathcal{K}$.
At each iteration, we solve problem \eqref{lp} (with the objective function $\hat{g}(\boldsymbol{\theta})$, $\boldsymbol{x}=\bar{\boldsymbol{x}}$, and $\boldsymbol{y}=\bar{\boldsymbol{y}}$) to obtain its solution $(\boldsymbol{z}^*, \boldsymbol{\theta}^*)$. 
If, for some service $k$, the E2E delay constraint is violated at this solution, we increase $\omega_k$ by a factor of $\rho > 1$, and solve problem \eqref{lp} again.
The procedure is repeated until the solution satisfies the E2E delay constraints of all services or the iteration number reaches a predefined parameter IterMax. 
We summarize the above procedure in Algorithm \ref{al:routing}.
\begin{algorithm}[h]
	\caption{{An iterative LP refinement procedure for solving the traffic routing subproblem}}
	\label{al:routing}
	\begin{algorithmic}[1]
		\renewcommand{\algorithmicrequire}{\textbf{Input:}}
		\renewcommand{\algorithmicensure}{\textbf{Output:}}
		{\STATE Set $\omega_k = 1$ for all $k \in \mathcal{K}$, $\rho >1$, $\text{IterMax} \geq 1$, and $t=0$;
		\WHILE {$t < \text{IterMax}$}
		\STATE Solve problem \eqref{mip} (with the objective function $\hat{g}(\boldsymbol{\theta})$, $\boldsymbol{x}=\bar{\boldsymbol{x}}$, and $\boldsymbol{y}=\bar{\boldsymbol{y}}$) to obtain its solution $(\boldsymbol{z}^*, \boldsymbol{\theta}^*)$;
		\STATE For each service $k\in \mathcal{K}$, compute the total delay $\bar{\theta}(k)$ based on $(\boldsymbol{z}^*, \boldsymbol{\theta}^*)$ and  $(\bar{\boldsymbol{x}}, \bar{\boldsymbol{y}})$.
		If $\bar{\theta}(k) \leq \Theta_k$ for all $k \in \mathcal{K}$, we stop with the feasible solution $(\boldsymbol{z}^*, \boldsymbol{\theta}^*)$; otherwise, for each $k\in \mathcal{K}$ with $\bar{\theta}(k) > \Theta_k$, we update $\omega_k \leftarrow \rho \omega_k$. Set $t\leftarrow t+ 1$.
		\ENDWHILE
	}
	\end{algorithmic} 
\end{algorithm}

{\bf\noindent$\bullet$ Complexity Analysis\vspace{0.1cm}}

The dominant computational cost of our algorithm is to solve the LP problems in form of \eqref{lp}. The number of solving problems \eqref{lp} in Algorithms \ref{roundingx} and \ref{al:routing} are upper bounded by $|\mathcal{V}|\sum_{k \in \mathcal{K}}\ell_k$ and $\text{IterMax}$, respectively.
Since an LP can be solved using the (polynomial-time) interior-point method \cite{Renegar1988}, it follows that the worst-case complexity of our proposed algorithm is polynomial.
%
%
%
In sharp contrast, the worst-case complexity of using the standard MILP solvers like Gurobi \cite{Gurobi} to solve problem \eqref{mip} is exponential.

\section{Numerical Simulation}

In this section, we present simulation results to illustrate the effectiveness and efficiency of our proposed LP rounding-and-refinement (\LPRR) algorithm for solving the network slicing problem.
We compare our proposed algorithm with the LP rounding (LPR) algorithm in \cite{Chowdhury2012} and the exact approach using standard MILP solvers (called EXACT) in \cite{Chen2020}. 
We choose $\sigma  = 0.001$ and $ |\mathcal{P}|=2 $ in problem \eqref{mip}.
In Algorithm \ref{al:routing}, we set $\rho = 2$ and $\text{IterMax}=5$.
We use Gurobi 9.0.1 \cite{Gurobi} to solve all MILP and LP problems.
When solving the MILP problems, we set a time limit of 1800 seconds for Gurobi. 

We test all algorithms on the fish network topology \cite{Zhang2017}, which contains 112 nodes and 440 links, including 6 cloud nodes.
%
%
The cloud nodes' and links' capacities are randomly generated within $ [50,100] $ and $ [5,55] $, respectively.
The NFV	 and communication delays on the cloud nodes and links are randomly generated within $\{3,4,5,6\}$ and $\{1,2\}$, respectively.
For each service $k$, node $S(k)$ is randomly chosen from the available nodes and node $D(k)$ is set to be the common destination node; SFC $ \mathcal{F}(k) $ is a sequence of functions randomly generated from $ \{f^1, \ldots, f^4\} $ with $ |\mathcal{F}(k)|=3 $; $ \lambda_s(k) $'s are the service function rates which are all set to be the same integer value, randomly generated within $ [1,11] $; $ \Theta_k $ is set to $ 20+(3*\text{dist}_k+\alpha) $ where $ \text{dist}_k $ is the delay of the shortest path between nodes $ S(k) $ and $ D(k) $ and $ \alpha $ is randomly chosen in $[0,5]$.
The above parameters are carefully chosen to make sure that the constraints in problem \eqref{mip} are neither too tight nor too loose. 
For each fixed number of services, 100 problem instances are randomly generated and the results presented below are obtained by averaging over these problem instances.

Fig.~\ref{figure1} plots the performance of \LPRR, LPR, and EXACT.
We can clearly see the effectiveness of our proposed algorithm \LPRR~over LPR in Figs.~\ref{nfeas} and \ref{sumy}.
In particular, as shown in Fig.~\ref{nfeas}, using our proposed algorithm \LPRR, we can find feasible solutions for much more problem instances, compared with using LPR.
Indeed, \LPRR~finds feasible solutions for almost all feasible problem instances (as EXACT is able to find feasible solutions for all feasible problem instances and the difference of the number of feasible problem instances solved by EXACT and \LPRR~is small in Fig.~\ref{nfeas}).
In addition, using \LPRR, the number of activated cloud nodes is much smaller than that of using LRP, as shown in Fig.~\ref{sumy}. 

The comparison of the solution efficiency of \LPRR, LPR, and EXACT is plot in Fig.~\ref{timeratio}. 
Here we scale the solution time of LPR to be 1 and compute the CPU time ratio as follows:
\begin{equation*}
\text{CPU time ratio} = \frac{\text{T(\LPRR)~(or T(EXACT))}}{\text{T(LPR)}},
\end{equation*}
where $\text{T(LPR)}$, $\text{T(\LPRR)}$, and $\text{T(EXACT)}$ are the CPU time taken by LPR, \LPRR, and EXACT, respectively.
Fig.~\ref{timeratio} shows that our proposed algorithm \LPRR~is much more computationally efficient than EXACT, and the solution efficiency of \LPRR~and LPR is comparable.
Indeed, \LPRR~is at most four times slower than LPR in all cases while EXACT is even 100+ times slower than LPR when the problem is large (i.e., $|\mathcal{K}| \geq 5$).

In summary, our simulation results illustrate the effectiveness and efficiency of our proposed algorithm \LPRR. More specifically, compared with LPR in \cite{Chowdhury2012}, it is able to find a much better solution; compared with EXACT in \cite{Chen2020}, it is much more computationally efficient.

\newpage

%

{}

}

\end{document}